\documentclass[%
reprint,
 amsmath,amssymb, 
 longbibliography, 
 aps,prl]{revtex4-1}

\usepackage{float}
\usepackage{graphicx}
\usepackage{dcolumn}
\usepackage{bm}
\usepackage[flushleft]{threeparttable}
\usepackage{url}
\usepackage{amsmath}
\usepackage{amssymb}
\usepackage{braket}
\usepackage{multirow}
\usepackage{array}
\usepackage{url}
\newcolumntype{P}[1]{>{\centering\arraybackslash}p{#1}}
\usepackage{xcolor}
\usepackage{framed}
\usepackage{lipsum}
\colorlet{shadecolor}{blue!60}
\usepackage{soul}
\usepackage{hyperref}
\usepackage{upgreek}
\usepackage{longtable}
\usepackage{ulem}
\makeatletter

\begin{document}

\preprint{APS/123-QED}

\title{Atomic-scale theory of robust out-of-plane ferroelectricity in ultrathin films}

\author{Fengbo Yuan$^{1}$, Yujia Teng$^2$, Karin M. Rabe$^2$, and Yubo Qi$^{1}$}

\affiliation{%
$^1$ Department of Physics, University of Alabama at Birmingham, Birmingham, Alabama 35233, USA \\
$^2$Department of Physics $\&$ Astronomy, Rutgers University, 
Piscataway, New Jersey 08854, United States 
}%

\date{\today}

\begin{abstract}

Ferroelectricity in ultrathin films, characterized by robust  switchable out-of-plane polarization, is key to next-generation nanoelectronics. 
Although the macroscopic theory of ferroelectricity suggests that ferroelectricity is inevitably suppressed as the film thickness decreases,  
recent studies have demonstrated robust ultra thin-film ferroelectricity, for certain ferroelectric materials, specifically HfO$_2$-based oxides and bismuth-based oxides.
In this work, we develop an atomic-scale theoretical framework for understanding ferroelectricity in this limiting regime.
By considering the work function of the termination layers of the film, we find that robust ferroelectricity arises from ``self-polarizing'' and ``switchable role of the termination layer'' effects strongly correlated to the ``characteristic structure.'' 
This theory also provides further insights on the importance of top electrodes in stabilizing ferroelectricity for this class of materials in the ultrathin limit.
This work aims to develop a comprehensive theoretical framework for thin-film ferroelectricity, providing fundamental insights that can guide the design of next-generation nanoscale devices.

\end{abstract}

\pacs{Valid PACS appear here}

\maketitle

Ferroelectricity, the property of possessing an electric-field-switchable polarization, 
plays an important role in next-generation devices, including non-volatile memory~\cite{Guo13p1990,Pradhan24p348}, nanoscale sensors~\cite{Yao22p2103842}, and field-effect transistors~\cite{Khan20p588,Mulaosmanovic21p502002}. 
However, achieving robust switchable polarization in the out-of-plane direction in ultrathin ferroelectric films has long been considered challenging~\cite{Batra73p3257,Mehta73p3379,Setter06p051606,Park23p2200096,Junquera03p506}.

The difficulty of maintaining a bulk-like polarization normal to the thin-film surface is generally understood to be the inevitable result of classical electrostatic considerations.
A free-standing polar slab, corresponding to a Type-3 surface in Tasker's classification~\cite{Tasker79p4977}, 
has surfaces with net charge  
generating a macroscopic electric field, called the depolarization field, which opposes the polarization~\cite{Dawber05p1083,Junquera03p506}.
The depolarization field is proportional to the polarization and thus does not in principle depend on thickness. 
Typically, screening of the depolarization field naturally occurs by adsorption of free charged species or by putting the slab between electrodes.
However, finite screening lengths lead to incomplete screening in thin films, 
resulting in a residual depolarization field that suppresses or can even still eliminate the ferroelectric polarization~\cite{Batra73p3257,Mehta73p3379,Wurfel73p5126,Kim05p237602,Qi15p044014,Zhang14p224101}.
This incomplete screening gets worse as the thickness of the film decreases.

Recent breakthroughs have revealed several unconventional ferroelectrics, most notably HfO$_2$-based systems, in which out-of-plane bulk-like spontaneous polarization is sustained in ultrathin films~\cite{Park18p795,Park15p1811, Boscke11p102903,Boscke11p112904,Muller11p114113,Xiao18p227601,Zhao20p12522,Yang23p1218}. 
Several seminal and emerging models have been proposed to explain the unconventional robustness of ultrathin ferroelectricity. 
In the improper ferroelectric models, the polar mode is coupled to other primary structural modes, such as lattice tilting, that remain stable in thin films, therefore protecting the polar mode from thickness-dependent suppression~\cite{Benedek11107204,Vogel23p18482,Sai09p107601,Zhou22peadd5953}. 
In the flat-band model, the flat polar band of HfO$_2$ indicates that the polarization of an individual unit cell is weakly affected by its neighbors, leading to the scale-free behavior~\cite{Lee20p1343}. 

Here, we would like to point out that the polar state of a ferroelectric thin film can be stabilized by the chemistry at the interfaces or surfaces, 
in a way that is quite different from screening of the depolarization field by the charges in electrodes. 
An adsorbate layer can modify the work function of the surface and help stabilize polarization.
For example, in addition to its electrostatic effect, an OH adsorbate can induce polarization pointing toward the surface~\cite{Fong06p127601,Bristowe12024106p,Copie17p29311,Yue19p245432,Spanier06p735}.  
Here, the adsorbate layer inducing stable polarization is referred to as ``polarizing layer.''
Besides adsorbates~\cite{Fong06p127601,Bristowe12024106p,Copie17p29311,Yue19p245432}, a surface reconstruction~\cite{Wang09p047601,Sepliarsky05p014110} or another material~\cite{Miao16p19965,Kolpak10p217601,Kumah10p251902} can also work as a polarizing layer.

In this work, we develop a first-principles theoretical framework that systematically includes the role of surfaces, interfaces and adsorbates in stabilizing polar states through a combination of electrostatic and work function effects.
This theoretical framework contains three principles related to polarizing layers.
Based on this theory, we show that the robust and switchable polarization in thin-film HfO$_2$ (and certain other ferroelectrics) can be understood recognizing that in a certain type of structure, which we call ``characteristic structures,'' the termination layers can act as polarizing layers, depending on the orientation of the dipolar distortion, leading us to describe these systems as ``self polarizing.'' 
This theoretical framework can account for stabilities of HfO$_2$ thin films with different oxygen coverages and why conventional ferroelectrics, such as PbTiO$_3$, can lose ferroelectricity in thin films.
These results provides practical guidelines for identifying new thin-film ferroelectrics.

\begin{figure}[hpbt]
\includegraphics[width=8.25cm]{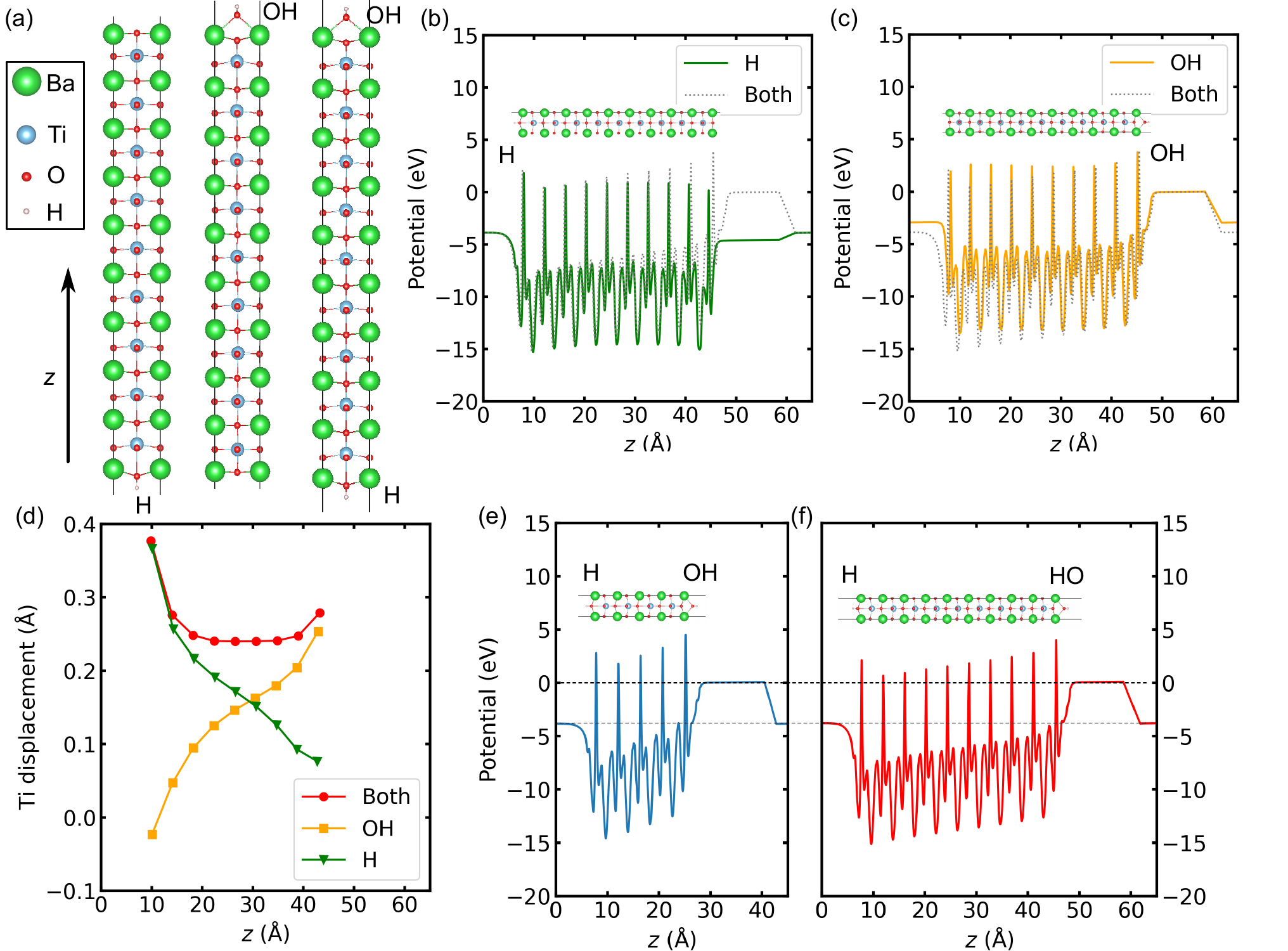}
\caption{(a) Structure of the BaTiO$_3$ thin films with OH, H, or both as adsorbates.
Macroscopic average potentials of BaTiO$_3$ thin films with (b) H and (c) OH adsorbates respectively.
(d) Ti displacements at different $z$-coordinates.
Macroscopic average potentials of (e) 1.7 nm and (f) 3.8 nm BaTiO$_3$ thin films.
}
\label{f1}
\end{figure}

\begin{figure}[hpbt]
\includegraphics[width=8.25cm]{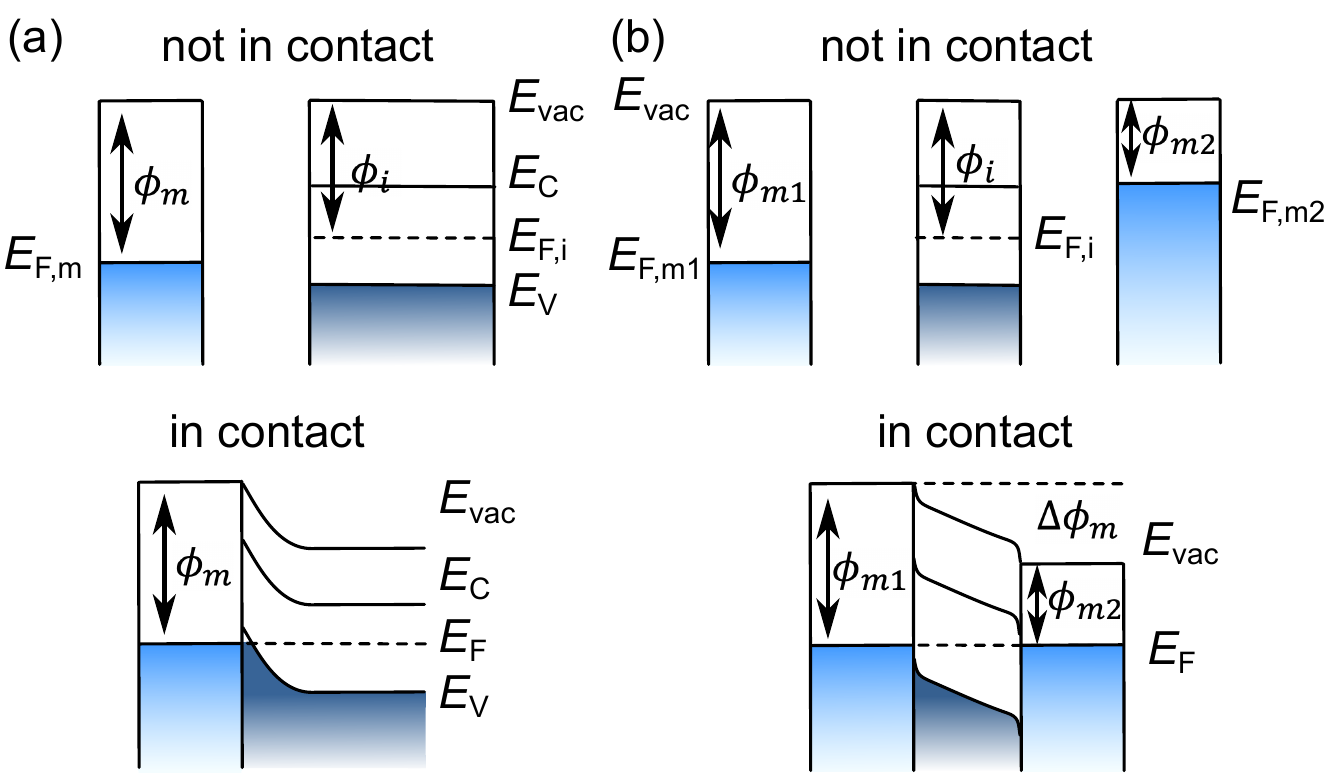}
\caption{Schematic plots of band alignments in (a) metal-insulator and (b) metal-insulator-metal heterostructures.
}
\label{f2}
\end{figure}

We begin with the ``single-sided polarizing layer'' principle, which says that a polarizing layer on one side can stabilize polarization in a thin film.
The system can be analogized to a metal-insulator heterostructure, as shown in Fig.~\ref{f2} (a).
After a metal is put in contact with an insulator with a different work function, electrons will redistribute, aligning the two Fermi levels, shifting the vacuum levels, and causing a bending of the bands in the insulator~\cite{Zhang12p5520}.
The polarizing layer will induce local polar distortion near the interface due to the band bending and associated electric field. 
This polar distortion can propagate into the film, depending on the coherence length for the distortion. 
As shown in Fig.~\ref{f1} (a), we consider BaTiO$_3$ thin films with OH, H, or both as adsorbates as illustrative examples [see Supplementary Materials (SM)~\cite{SM} section I for computational details].

By comparing results with those for systems with adsorbate layers on both sides, we can see that these effects of the polarizing layer are quite local and transferable. 
Specifically, in Fig.~\ref{f1} (b) and (c), we plot the macroscopic average potentials of the BaTiO$_3$ thin film with OH adsorbates and H adsorbates at only one side respectively and compare them with the potential profile of the thin film with adsorbates at the both side.
We find that the macroscopic average potential profiles near the adsorbates are approximately identical for single-sided and double-sided adsorption.
For the single-sided polarizing layer case, the potential profiles are approximately flat deep in the films, and the Ti displacements decay with increasing distance from the polarizing layer [Fig.~\ref{f1} (d)].

The next principle is the ``thickness independent potential'' principle, which says that if there are polarizing layers at both sides of a thin film, the difference in the potential in the vacuum on the two sides is thickness independent, being the difference in work function of the two adsorbate layers.
Here, we consider a BaTiO$_3$ thin film with OH and H adsorbates on both sides.
Fig.~\ref{f1} (e) and (f) show the macroscopic average potentials of BaTiO$_3$ slab with different thicknesses.
We observe that the vacuum potential difference between the two sides is a constant regardless of the thickness of the slab.
We can understand this by regarding the adsorbate layers as metal layers in a metal-insulator-metal heterostructure. 
As shown in Fig.~\ref{f2} (b), if an insulator is sandwiched between metals with different work functions, 
charge will transfer directly from one metal layer to the other and the potential change through the insulator should be $\Delta\phi$, which is the difference of work function between the two metals~\cite{Zhang12p5520}.
Such a principle is general in thin films with polarizing layers at both sides, though if the charge transfer is large, it might slightly change the work function of the polarizing layer (See SM section II for more examples and exceptions of this principle).
An important implication of ``thickness independent potential'' principle is that a thinner film has a larger internal electric field.
A thin film may become unstable due to the large electrostatic energy, which is related to the ``over polarization'' effect discussed next.

The ``thickness independent potential'' principle demonstrates that the different work functions of the two polarizing layers induce a potential difference and can stabilize polarization in the film.
In such a case, the electric field in the film is determined by the work function difference and also the thickness of the film.
If the electric field is too large, the polar phase of the thin film may become unstable, which is referred as to the ``over polarization'' principle.
Over polarizing typically occurs when the two polarizing layers are composed of dense, highly ionic, and oppositely charged ions.
Two types of structural transformation can occur under over polarizing.
In the first type, the entire slab, including the polarizing layers and the film interior, transforms into a non-polar structure.
In the second type, the polarizing layers may change their bonding patterns, to reduce the work function difference and internal electric field.

So far, we have reviewed that polarizing layers can stabilize polarization but over polarizing can induce instability.
However, most polar slabs stabilized by polarizing layers cannot be viewed as ferroelectric,
since a preferred polarization direction has been predetermined by the work functions of the polarizing layers, which is strongly related to the electronegativity (See SM section III for more details).
Applying and then removing an electric field cannot switch the polarization direction.
Polarization switching is expected to be associated with complex processes such as adsorbate detachment or adsorbate penetrating through the slab.

However, it has been observed that thin films of some ferroelectrics, most notably HfO$_2$, can exhibit a robust switchable polarization down to the unit cell scale, even without any adsorbate layers or electrodes. Here, we show that this can be explained by the fact that for certain crystal structures, which we call ``characteristic structures.'' the termination layers of the film can act as two different types of polarizing layers, switching between types with the polarization direction, in a mechanism that we refer to as ``self polarization.'' 

\begin{figure}[hpbt]
\includegraphics[width=8.25cm]{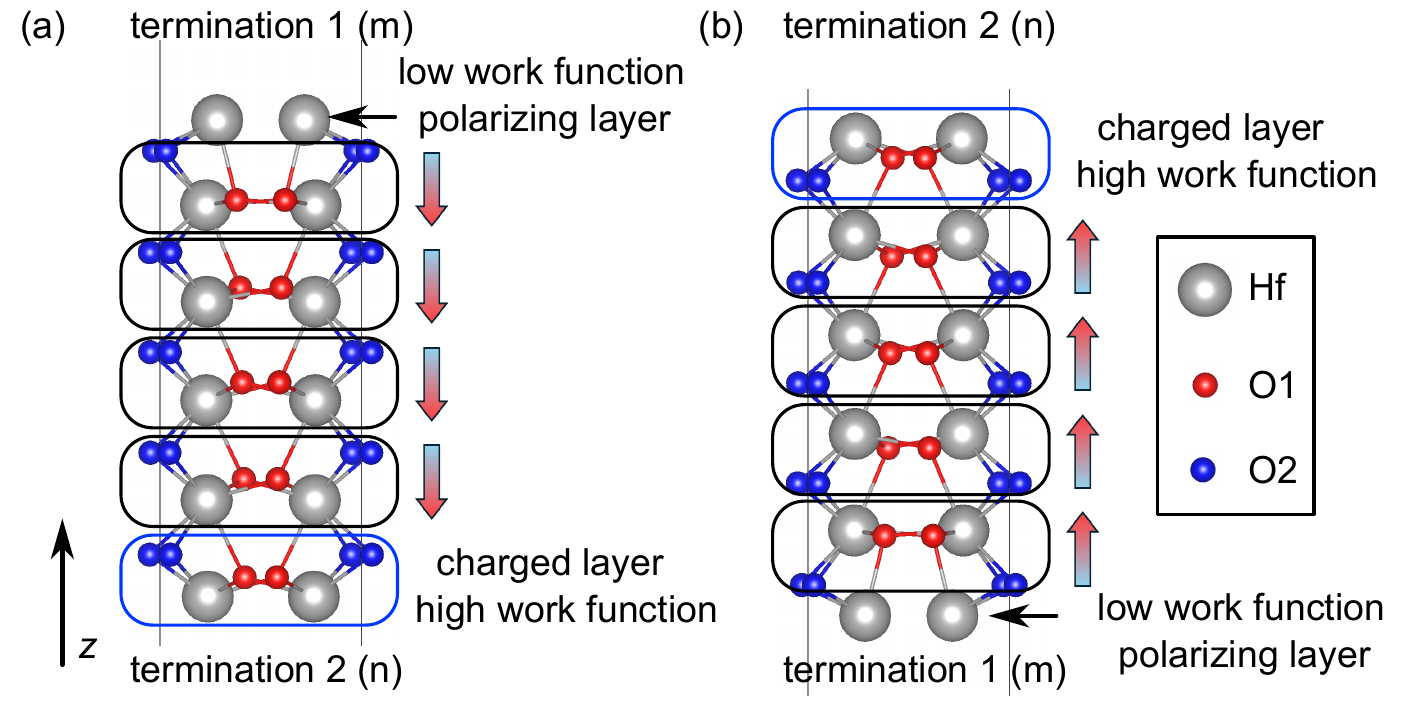}
\caption{(a) and (b), optimized structures of 6-Hf-layer HfO$_2$ with Hf terminations at both ends.}
\label{f3}
\end{figure}

We start with a HfO$_2$ slab with Hf terminations on both sides, as shown in Fig.~\ref{f3}, and extend the discussion to other terminations later. 
In the ferroelectric $Pca2_1$ structure of HfO$_2$~\cite{Sang15p162905}, there are two categories of oxygen atoms: three-coordinated (O1) and four-coordinated (O2)~\cite{Yu25p114219}. 
The O1 atoms locate near the mid-plane of their adjacent Hf layers, while the O2 atoms are displaced from the mid-plane toward Hf layers, leading to non-zero polarization.

The slab structures presented in Fig.~\ref{f3} (a) and (b) are fully relaxed using DFT calculations and preserve their polar $Pca2_1$ structures. 
For the structure in Fig.~\ref{f3} (a), the bottom layer of Hf atoms bond closely to O1 atoms, while the top layer does not. 
We group a layer of Hf atoms together with the O1 and O2 layers above it together and approximate it as a dipole (since the O1 above bonds strongly to Hf).
On the other hand, the top Hf layer, which bonds with O1 weakly, is approximated as a metal with a small electronegativity, serving as a low-work-function polarizing layer.
The top Hf layer can stabilize a polarization pointing away from it, similar to the H adsorbate on a BaTiO$_3$ slab.
The top layer has excessive free conducing electrons and transfer some to the bottom layer, which serves as a high-work-function polarizing layer, since the Hf atoms at bottom are also underbonded, as shown in Fig. S3. 
The top and bottom polarizing layers together stabilize the polarization in the thin film.
In Fig.~\ref{f3} (b), the resultant structure after polarization switching, the top layer has switched to the high-work-function polarizing layer structure, and the bottom layer has switched to the low-work-function polarizing layer structure.  
Thus, we attribute the switchable polarization in HfO$_2$ thin film to the switchability of the termination layers between the low-work-function and high-work-function polarizing layers structures.

\begin{figure}[hpbt]
\includegraphics[width=8.25cm]{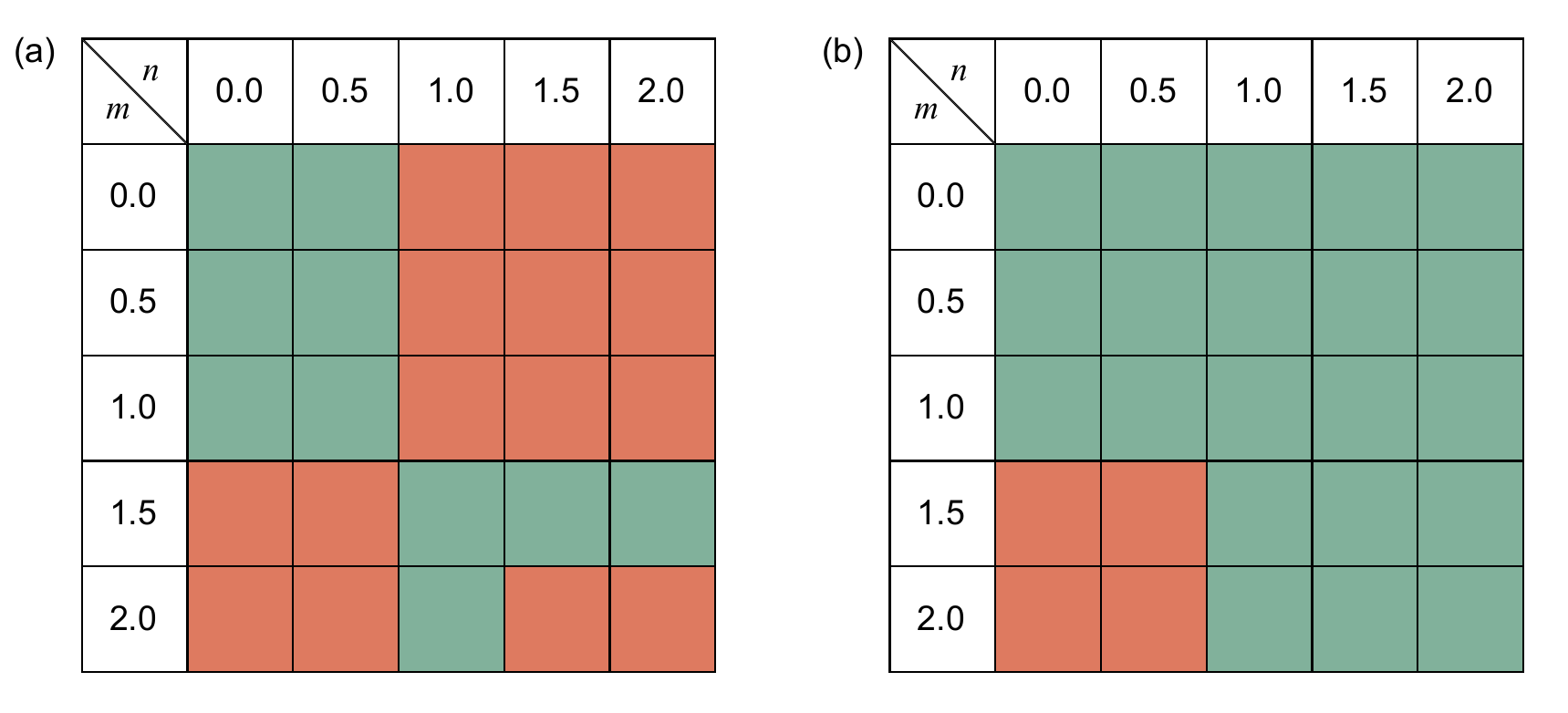}
\caption{Stabilities of (a) 6-Hf-layer and (b) 10-Hf-layer HfO$_2$ slabs with different oxygen coverages at two terminations. Green and red colors indicate stable and unstable structures respectively (see SM sections IV and V for the relaxed structures).
}
\label{f4}
\end{figure}

Next, we extend our discussion to the stabilities of HfO$_2$ slabs with other terminations. 
We refer to the termination whose Hf atoms are far away from O1 as termination 1, and the other termination as termination 2. 
Following the notations in Refs.~\cite{Acosta21p124417,Acosta23p124401}, 
we use $m/n$ to represent the oxygen coverage, where $m$ and $n$ denote the number of oxygen atoms per surface formula unit on terminations 1 and 2 respectively. 
We carry out DFT relaxations on 6-Hf-layer HfO$_2$ polar slabs with different coverage of oxygen atoms, 
and their stabilities are summarized in Fig.~\ref{f4} (a). 
Here, if the slab remains the ferroelectric polar structure after relaxation, it is considered stable. 
We do not consider whether there are competing phases with lower energy, since a metastable polar state means that the polarization can sustain after the removal of electric field. 
We note, however, that the polarization of a given $m/n$ state may switch spontaneously on relaxation to its oppositely polarized structure $n/m$ if it is unstable.

Next, we study the effect of changing the work functions of one or both polarizing layers. 
By adding oxygen atoms to a given termination, we increase the work function. 
In Fig.~\ref{f4} (a), we see that increasing the work function by increasing the number of oxygen atoms $m$ on the top surface layer in Fig.~\ref{f3} (a) tends to destabilize the polar state, since the single-sided polarizing effect of the layer is decreased and the work function difference between the top and bottom layers is decreased. 
Increasing the number of oxygen atoms $n$ on the bottom layer increases the work function difference, initially further stabilizing the polar state, but subsequently destabilizing the polar structure by over polarizing. 
Increasing $m$ and $n$ together reduces the change in work function difference, promoting stability of the polar state. 
For the largest values of $m$ and $n$ (the 2.0/1.5 and 2.0/2.0 cases), instability is produced by an unfavorable energetics for the high concentration of oxygen atoms at the surface. 
Depending on $m$, $n$ and the thickness, instability of the polar state could be to a nonpolar state (Fig. S5 and Fig. S6) or to a polar state with opposite polarization (Fig. S10).

Next, we study the effect of increasing film thickness by carrying out DFT relaxations on 10-Hf-layer HfO$_2$ polar slabs.
The mechanism described above by which the addition of oxygens to the top layer tends to destabilize the polar film is thickness independent, and indeed the $m$/0.0 and $m$/0.5 films are unstable for $m$ greater than or equal to 1.5, just as in the 6 layer case [bottom left corner of Fig.~\ref{f4} (b)]. 
On the other hand, the addition of oxygens to the bottom layer tends to destabilize the polar state of the film by over polarizing above a critical value of $n$. 
The thickness independence of the potential difference makes the electric field smaller for thicker films, raising the critical value of $n$ (see SM section V for the macroscopic average potential profiles).
Indeed, the terminations that were unstable for 6-layer films are stable for 10 layer films [top right corner of Fig.~\ref{f4} (b)]. 
For the largest values of $m$ and $n$ (the 2.0/1.5 and 2.0/2.0 cases), the structures are stable for the 10-layer films because the surface energy has less influence on thicker systems.

The role of electrodes can also be understood from our work function model. 
When electrodes are added at both ends, free charge is redistributed among the two electrodes and the two polarizing layers to equalize the Fermi level. 
One result is that the difference in the vacuum potential is determined by the difference in the work function of the electrodes [see Fig. S9 (a) in SM section VI] and is zero for symmetric electrodes. The electric field in the film will also change when electrodes are added.

\begin{figure}[hpbt]
\includegraphics[width=8.25cm]{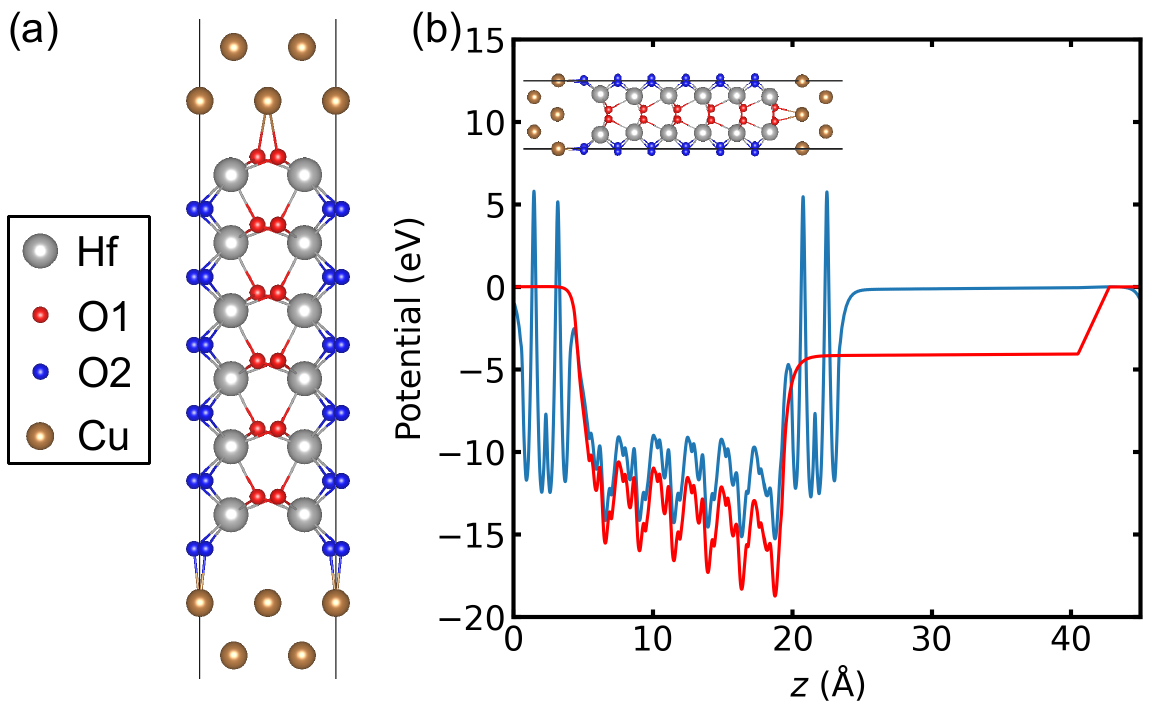}
\caption{Comparison between the macroscopic average potentials of a 6-Hf-layer HfO$_2$ polar slab with Cu electrodes (blue curve) and the same slab without Cu electrodes (red curve). 
}
\label{f6}
\end{figure}

We carry out DFT calculations on 6-Hf-layer slabs with copper electrodes.
In Fig.~\ref{f6} (a), we show the fully relaxed 1.0/1.0 6-Hf-layer slab between Cu electrodes as an illustrative example. 
We compare its macroscopic average potential with the potential computed with the electrodes removed and the other atoms fixed in position (this structure is unstable under full relaxation). 
As shown in Fig.~\ref{f6} (b), the electric field inside the film decreases compared with the free-standing case.
These results provide additional insights on the well-established experimental results that the addition of electrodes plays a critical role in stabilizing the ferroelectric phase~\cite{Fields22p2200601,Karbasian17p022907,Feng24p015105,Cao18p1207}. 
As shown in Fig. S9 (b) and (c), electrodes at two sides can stabilize 6-Hf-layer HfO$_2$ slabs with a wide range of O coverages, but electrode at only one end cannot (see SM section VI for the relaxed structure with Cu electrodes). 
Recognizing that the addition of a top electrode can have multiple effects, such as mechanical confinement~\cite{Bai23p174102,Wang23p15657}, 
our results are consistent with our proposal that the internal electric field reduction plays an important role. 

Our results attribute the existence of robust ferroelectricity in HfO$_2$-based thin films to the following two key effects.
First, the polarizing layer is part of the material itself, not requiring external adsorbates, and such an effect is referred to as ``self-polarization''. 
Moreover, a termination Hf layer (and oxygen atom on it) can be either a low-work-function polarizing layer (similar to a metal) or a charged layer under different polarization directions, and such an effect is referred to as ``switchable role of the termination.''

We also notice that these two key effects are the consequence of the crystal structure. 
We refer to such a family of structure as a ``characteristic structure'', as shown schematically in Fig. S12 (a) and (b). 
In a ``characteristic structure'', the slab has the same atoms on both terminations, but the two termination layers bond to the adjacent atomic layers differently, so that one termination acts as a low work-function polarizing layer and the other as a high work-function polarizing layer. Then, when the polarization switches, the characters of the two polarizing layers switch as well.
Many other recently discovered thin film ferroelectrics, including the recently discovered 1-nm ferroelectric bismuth oxide~\cite{Yang23p1218} and theoretically proposed (111)-oriented PbTiO$_3$ films~\cite{Zhu25p12952}, adopt ``characteristic structures'' and are expected to have robust ferroelectricity in ultrathin films by our theoretical model (see SM section VII for their structures).

On the other hand, in (001)-oriented conventional ABO$_3$ (such as BaTiO$_3$ or PbTiO$_3$) symmetrically-terminated perovskite ferroelectric thin films,  
the termination layer (either AO or BO$_2$) is charge neutral, having a medium electronegativity,  and is unlikely to work as a polarizing layer. 
Surface reconstruction with vacancies or adatoms can make the surface layer non-neutral and serve as polarizing layer.
But the polarization has a preferred direction and cannot be switched by the application and then removal of an electric field.


In summary, we develop a theoretical framework explaining the observation of robust ferroelectricity in ultrathin films ferroelectricity from the work function point of view. 
The theory explains the robust switchable polarization as a consequence of the ferroelectrics having a ``characteristic structure.''
In a ``characteristic structure'', a termination layers of the material act as polarizing layer whose role can switch under different polarization direction, enabling a switchable and robust polarization through the thin film. 
The theory, consisting of three principles, provides further insight into the importance of top electrodes in stabilizing the HfO$_2$ ferrolectric phase.
This work serves as a comprehensive theoretical framework for ferroelectricity in ultrathin films and provides guidance on the design of next-generation nanoscale device.

\section*{acknowledgement}
F.Y. and Y. Q. acknowledge support from the National Science Foundation under Grant No. OIA-2428751.
Y.T. and K.M.R. acknowledge support from the Wisconsin Materials Research Science and Engineering Center (NSF DMR-2309000), and the Office of Naval Research through N00014-21-1-2107.
First-principles calculations were performed using the computational resources provided by the Cheaha Supercomputer at the University of Alabama at Birmingham, and the High-Performance Computing Modernization Office of the Department of Defense.

\bibliography{cite}
\end{document}